\documentclass[a4paper]{article}

\usepackage{17lomcon}        
\usepackage{cite}             
\usepackage{epsfig}           

\usepackage{graphicx}
\usepackage{caption}
\usepackage{caption}

\usepackage{subcaption}
\usepackage{wrapfig}
\usepackage{lineno,xcolor}

\begin{document}


\title{Top physics in ATLAS }

\author{Roger Naranjo \email{roger.naranjo@desy.de} \\ on behalf of the ATLAS collaboration}

\affiliation{DESY, Hamburg, Germany.}


\date{}
\maketitle


\begin{abstract}
   These proceedings summarize the latest measurements on top production, top properties and searches using the ATLAS detector at the LHC. The measurements are performed on $pp$ collision data with a center of mass energy  $\sqrt{s} = 7, 8$ and  $13$ TeV.
\end{abstract}

\section{Introduction}

The top quark is the heaviest known elementary particle of the Standard Model (SM). It has a life-time shorter than the hadronization time and offers a window to search for new physics. With large numbers of top events produced at LHC in Run-1, it is possible to study its properties in detail. These proceedings present measurements of top production, top properties and searches at the center of mass energy of 7, 8 and 13 TeV by the ATLAS experiment~\cite{ATLAS}.

\section{Top production measurements}

\subsection{Top pair production}

\begin{figure}[h]
\centering

\begin{subfigure}[b]{0.50\textwidth}
        \includegraphics[width=\textwidth]{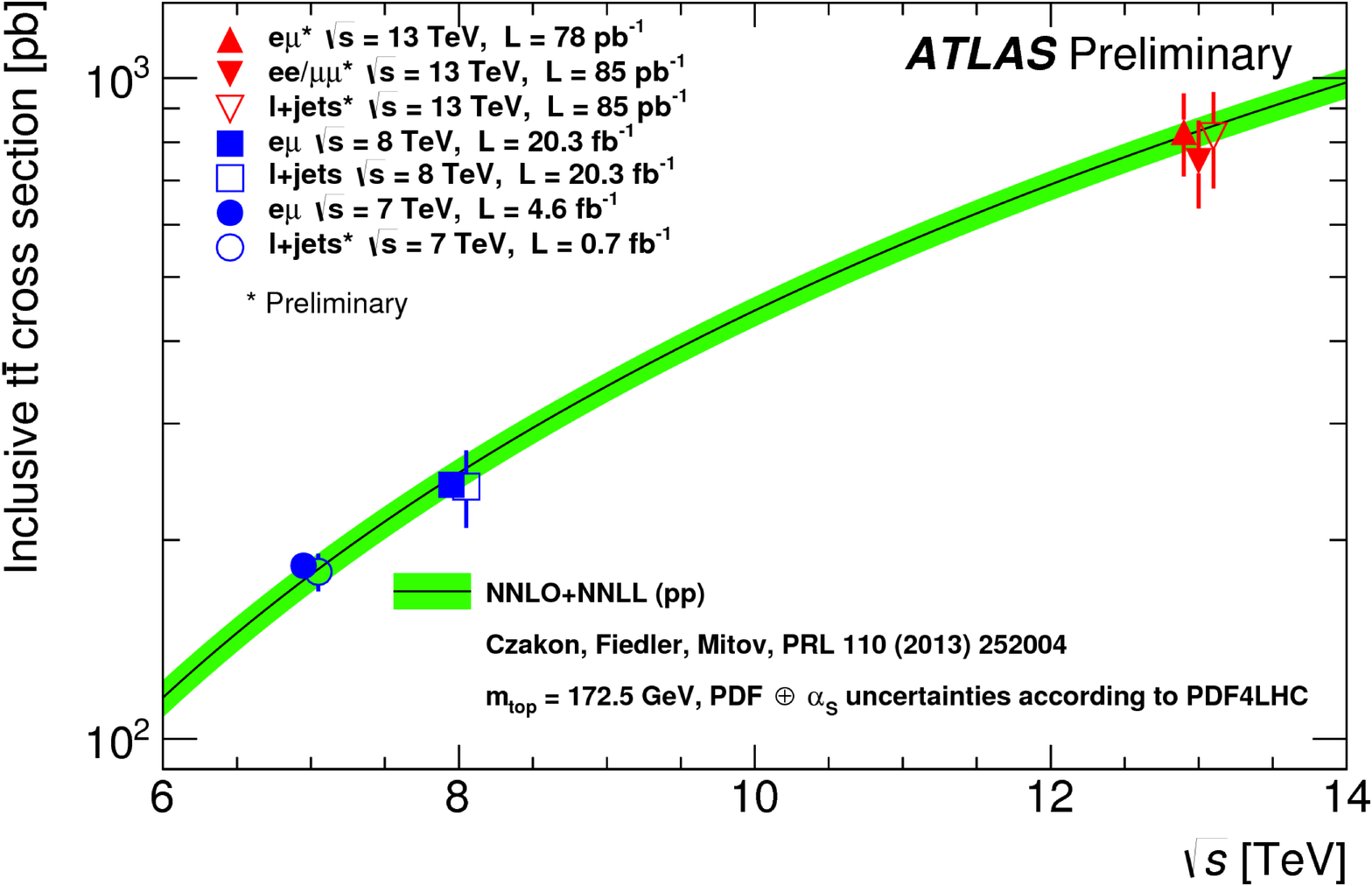}
        \caption{}
        \label{fig:a}

    \end{subfigure}
    \begin{subfigure}[b]{0.40\textwidth}
        \includegraphics[width=\textwidth]{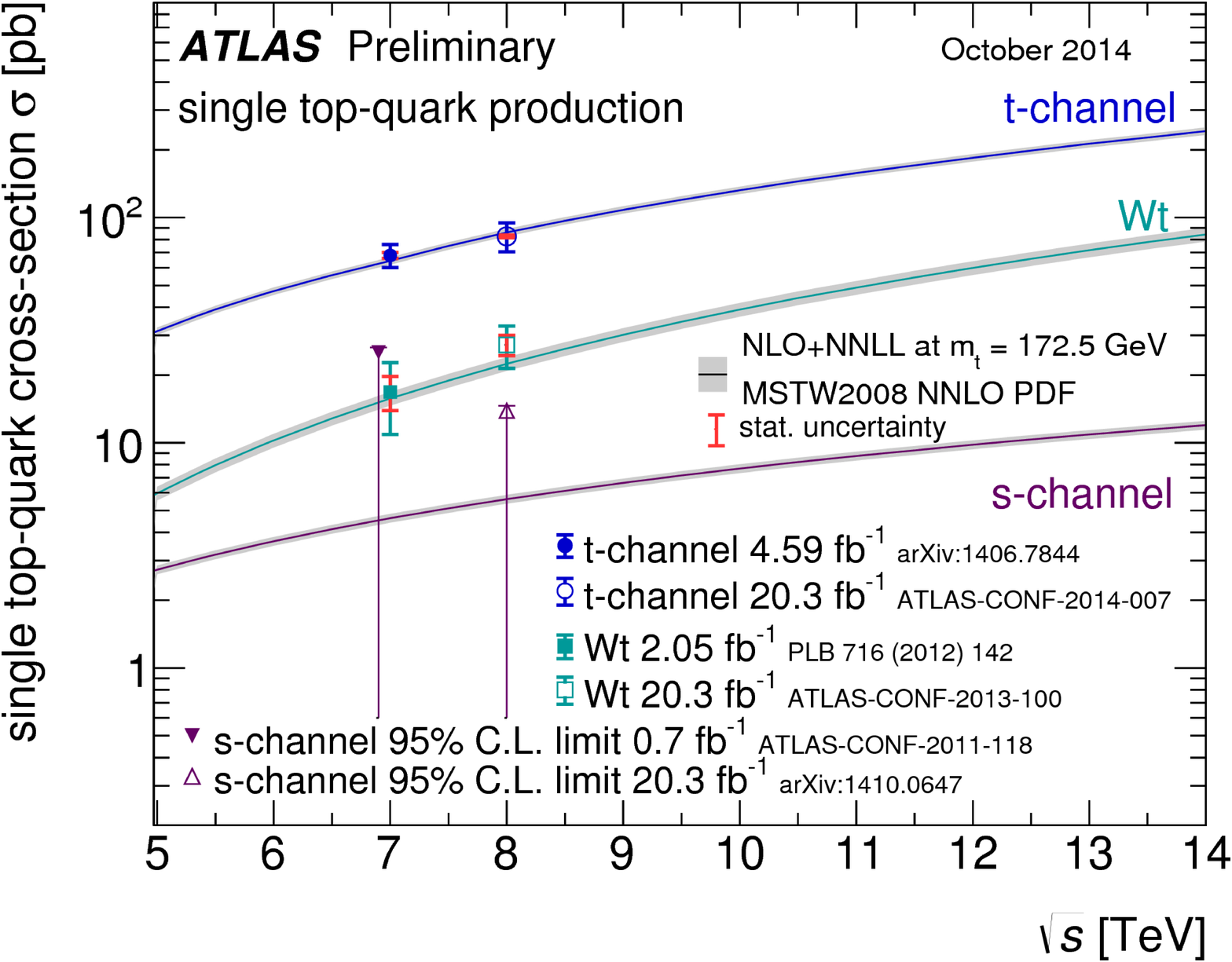}
        \caption{ }
        \label{fig:b}

    \end{subfigure}

\caption{(a) Cross section for $t\bar{t}$ pair production in pp collisions as a function of centre-of-mass energy. The results at $\sqrt{s}=13$, $8$ and $7$ TeV are compared to the theory predictions~\cite{Czakon:2013goa}. (b) Summary of ATLAS measurements of the single top production cross sections in various channels as a function of the center of mass energy compared to theoretical calculations~\cite{singletopSummary}.}
\end{figure}

The ATLAS experiment has performed measurements of the inclusive top-antitop production cross section of $t\bar{t}$ production at center of mass energy of 7, 8 and 13 TeV in the dilepton channel ~\cite{Aad:2014kva,crosssectiondilep}. 
The measurement at 8 TeV represents the
most precise measurement to date, even more precise than next-to-next-to-leading order (NNLO) predictions~\cite{Czakon:2013goa}. At 13 TeV, the cross section is found to be $\sigma_{t\bar{t}}= 825 \pm 49 \ ({\rm stat}) \pm 60 ({\rm syst}) \pm 83 ({\rm lumi})~{ \rm pb}$. In the three analyses, a simultaneous fit of the cross section, b-jet reconstruction and tagging efficiency is performed in the $e\mu$ channel. All the measurements are in agreement
with SM predictions, as shown in Figure~\ref{fig:a}.

New physics could affect the shape of the differential cross section as a function of top kinematics. In this conference, measurements performed in two different topologies were presented: the resolved and the boosted topologies, which are optimized for tops with transverse momentum $p_{\rm{T}} < 300$ and  and $p_{\rm{T}} > 300$ GeV respectively. In the resolved topology, the measurement of the top-antitop differential cross section is performed  as a function of the mass, $p_{\rm{T}}$ and rapidity of the top pair system at $7$ TeV~\cite{Aad:2015eia}. These measurements are performed in a fiducial region closely matching the detector acceptance and using tops observables based on stable particles in the $e\mu$ channel. 

\begin{wrapfigure}[16]{r}{0.4\textwidth}
\centering
\includegraphics[width=0.4\textwidth]{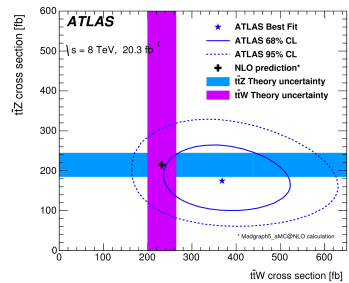}
\caption{The result of the simultaneous fit to the $t\bar{t}W$ and $t\bar{t}Z$ cross sections along with the $68\%$ and $95\%$ CL uncertainty contours~\cite{Aad:2015eua}. }
\label{fig:ttWttV}
\end{wrapfigure}
The observed spectra are softer than the Monte Carlo (MC) predictions. In the boosted topology, the differential cross section as a function of the $p_{\rm{T}}$ of the $t\bar{t}$ system is measured in the semileptonic decay channel at $8$ TeV~\cite{Aad:2015hna}. The measurements are performed in both a fiducial and total phase space. The measurements are also found to be softer than the MC predictions.

\subsection{Single top production}

Several measurements of the inclusive single top cross section in the $s$, $t$ and $Wt$ channel at 7 and 8 TeV have been performed by ATLAS. The latest 8 TeV results were presented in this conference. Figure~\ref{fig:b} shows a summary of these measurements~\cite{singletopSummary}. The inclusive cross sections for $t$-channel is found to be $\sigma_t = 82.6 \pm 1.2 \, (\rm{stat.}) \pm 11.4 \, (\rm{syst.}) \pm 3.1 \, (\rm{PDF}) \pm 2.3 \, (\rm{lumi.})$ pb, showing a good agreement with the NLO+NNLL\footnote{Next-to-leading order plus next-to-next-to-leading-logarithm} predictions~\cite{singletop1}, as well as the inclusive measurement in the $Wt$ channel where the measured value is $\sigma_t = 23.0 \pm 1.3 \, (\mathrm{stat.}) \pm {}^{+3.2}_{3.5} \, (\mathrm{syst.}) \pm 1.1 \, (\mathrm{lumi.})$ pb~\cite{singletop2}. An upper limit for the $s$-channel cross section was established~\cite{singletop3}.

\subsection{Top pair production in association with vector bosons}

The $t\bar{t}W$ and $t\bar{t}Z$ cross sections were measured in final states with two, three or four leptons~\cite{Aad:2015eua}. In this analysis, a simultaneous fit is performed in 20 signal and control regions. The control regions are defined by varying the number jets and lepton charge in the event. The results agree with the SM prediction as shown in Figure~\ref{fig:ttWttV}. 
\\

\section{Top quark properties}
\subsection{Mass}

Measurements of the top mass, performed in $t\bar{t}$ events in the semileptonic and the dileptonic channels, were presented~\cite{Aad:2015nba}. The analysis in the semileptonic channel uses a 3-dimensional template fit which determines simultaneously the top quark mass, the jet and $b$-jet energy scale factor. This measurement is the most precise mass measurement performed by ATLAS to date. The dilepton channel uses a one dimensional template fit and the combined result is found to be $m_{\mathrm top} = 172.99 \pm 0.48({\rm stat.}) \pm 0.78({\rm syst.})$ GeV.

In single top decays, the measurement is performed through a one dimensional template fit using the invariant
mass of the lepton and $b$-jet, in order to reduce the modelling uncertainties~\cite{ATLAS:2014baa}. Neural networks are used to optimize the
purity of the selection to $50\%$. The measured value is found to be $m_{\mathrm{top}} = 172.2 \pm 0.7 {\mathrm{(stat.)}} \pm 2.0 {\mathrm{(syst.) \, GeV}}$. 

Moreover, a measurement of the pole mass using the normalized differential cross section for $ t\bar{t} + 1$-jet as a function of the inverse of the invariant mass of system was performed~\cite{Aad:2015waa}. The selected events are identified using the lepton+jets top-quark pair decay channel, where lepton refers to either an electron or a muon. The measured value is found to be $m_{\mathrm{top}} = 173.7 \pm 1.5 {\mathrm{(stat.)}} \pm 1.4 {\mathrm{(syst.)}} {}^{+1.0}_{-0.5} {\mathrm{(theory) \, GeV}}$.

\subsection{Spin correlation}

Top pairs produced via the strong interaction are produced unpolarized at leading order. Their spins are correlated and information transferred to decay products. A measurement of the $t\bar{t}$ spin correlation  at $8$ TeV was performed~\cite{Aad:2014mfk}. In this analysis, the $t\bar{t}$ spin correlation is extracted from dilepton events by using the difference in the azimuthal angle of the two selected leptons. The measured value is consistent with the SM predictions. Moreover, the measurement of the angular distribution of the decay products is sensitive to super symmetric top squark (stop) pair production. The top squarks with masses between the top quark mass  172.5 GeV and 191 GeV are excluded at the $95\%$ confidence level.

\section{Searches}
In this conference, limits in searches for flavor changing neutral currents (FCNC) were presented~\cite{Aad:2015uza}. FCNC are predicted
by the SM but are largely suppressed by the GIM mechanism. In Beyond the Standard Model (BSM) models the FCNC are enhanced, giving a window to search for new physics. One search is performed in $t\bar{t}$ events with one top quark decaying to a $Z$-boson and a light quark and the other to a $W$-boson and a $b$-quark. Both bosons are required to decay
leptonically. No evidence of FCNC is found and an observed upper limit is established. Another search for FCNC is performed in single top production via $gu/c \rightarrow t$ with the top decaying to a $Z$-boson and a light quark~\cite{Aad:2015uza}. Only the leptonic channel is considered and upper limits are established on the cross section times branching ratio and on the coupling constants of the processes.

\section{Conclusions}
ATLAS has performed multiple measurements in the top physics domain. The measurements of cross sections and properties are in agreement with the SM predictions. Some top quark properties were used to exclude BSM models and limits were established for FCNC searches.


\end{document}